Adem Chanie Ali*, Seid Muhie Yimam, Abinew Ali Ayele, Chris Biemann and Martin Semmann

# Silenced voices: social media polarization and women's marginalization in peacebuilding during the Northern Ethiopia War

**Abstract:** This study examines the complex relationship between social media, polarization, and conflict, with a focus on digital peacebuilding and women's participation, using the Northern Ethiopia War as a case study. Using a qualitative exploratory design through in-depth interviews, focus groups, and document analysis, the research examines how social media platforms influence conflict dynamics. The study applies and advances social identity, liberal feminist, and intersectionality theories to analyze social media's role in shaping conflict, mobilizing ethnic politics, and influencing women's involvement in peacebuilding. Findings reveal that the weaponization of social media intensifies polarization and offline violence. Women are disproportionately impacted through displacement, exclusion from peace negotiations, and heightened risks of gender-based violence, including rape. Contributing factors include hostile online environments, the digital divide, and prevailing socio-cultural norms. The study identifies significant gaps in leveraging digital platforms for sustainable peace, including government-imposed internet shutdowns, unregulated social media environments, and low media literacy. It recommends media literacy initiatives, inclusive peacebuilding frameworks, open and safe digital spaces, and gender-sensitive technological approaches. By centering digital technology, conflict, and gender in the Global South, this research contributes valuable insights to ongoing debates on ICT in conflict, peacebuilding, and women's empowerment.

***Corresponding author: Adem Chanie Ali**, Department of Journalism and Communication, Bahir Dar University, Bahir Dar, Ethiopia; and Hub of Computing and Data Science, University of Hamburg, Hamburg, Germany, E-mail: adem.ali@uni-hamburg.de. https://orcid.org/0000-0002-7035-8951
**Seid Muhie Yimam**, **Chris Biemann** and **Martin Semmann**, Hub of Computing and Data Science, University of Hamburg, Hamburg, Germany, E-mail: seid.muhie.yimam@uni-hamburg.de (S.M. Yimam), chris.biemann@uni-hamburg.de (C. Biemann), martin.semmann@uni-hamburg.de (M. Semmann). https://orcid.org/0000-0002-8289-388X (S.M. Yimam). https://orcid.org/0000-0002-8449-9624 (C. Biemann). https://orcid.org/0000-0001-5316-3696 (M. Semmann)
**Abinew Ali Ayele**, ICT4D Research Center, Bahir Dar Institute of Technology, Bahir Dar University, Bahir Dar, Ethiopia, E-mail: abinew.ali.ayele@uni-hamburg.de. https://orcid.org/0000-0003-4686-5053

## 1 Introduction

The Northern Ethiopia War, which lasted from November 3, 2020, to November 3, 2022, primarily unfolded in the Tigray region and involved the Ethiopian Federal Government, Eritrea, and various regional forces. The conflict started when the Ethiopian government accused Tigray forces of attacking a military base, leading to a humanitarian crisis displacing over 20 million people, especially women and children.[1] During the two-year conflict, the Tigray region and the neighboring regions of Amhara and Afar experienced severe damage to essential social services, including the education sector, hospitals, industries, and other infrastructures. The conflict resulted in significant losses of life, with estimates of casualties ranging from 311,000 to 808,000, with an average estimate of 518,000.[1,2] Instances of war rape were reported to be frequent, with girls as young as 8 and women as old as 72 being subjected to sexual violence, often in front of their families.[2,3,3] The violent armed conflict came to a halt following the signing of peace agreements between the warring factions in Pretoria and Nairobi in November 2022.[4] The complexity of this conflict lies not only in its

1 Humanitarian Evaluation of the Northern Ethiopia Crisis.
2 The Guardian – Ethiopia's devastating war.
3 Rape as a War Crime.

immediate humanitarian crisis but also in its intricate web of deeply rooted ethnic tensions and struggles for political power.

In a volatile environment, social media platforms like Facebook, YouTube, X (Twitter), and Telegram have emerged as critical tools for communication and mobilization. These platforms influence socio-political issues such as conflict dynamics and social cohesion in both positive and negative ways.[5,6]

On one hand, social media can facilitate peacebuilding initiatives, support democracy movements, raise awareness, and empower marginalized groups.[5–7] Sokfa's[8] study highlights that the reliance on digital tools such as social media, mobile apps, and crowdsourcing platforms has become increasingly prominent in addressing and potentially mitigating conflict. These technologies present innovative opportunities for conflict prevention, mediation, and reconciliation, including early warning systems and platforms for dialogue facilitation. On the other hand, social media can also spread polarization, hate speech, and incite violence.[7,9] The same digital platforms that can foster peace also carry significant risks, such as the spread of misinformation and government surveillance, which can exacerbate existing social tensions.[8]

Thus, understanding the role of social media in the context of the Northern Ethiopia War is crucial for unpacking its impact on conflict dynamics and societal fragmentation, particularly in light of its implications for gender and social justice.

This study seeks to address a vital question: How do social media platforms influence the dynamics of polarization and conflict during the Northern Ethiopia War, particularly regarding women's participation in peacebuilding? To address this question, the research aims to achieve the following objectives:

1. Analyze how social media has contributed to polarization and conflict during the Northern Ethiopia War, with particular attention to its impact on women's experiences and roles.
2. Investigate the extent of women's participation in, and exclusion from, the peacebuilding processes that led to the resolution of the Northern Ethiopia War.
3. Highlight existing digital peacebuilding endeavors and identify gaps in these efforts.
4. Identify the challenges of digital peacebuilding, particularly concerning gender dynamics.

The study has significant theoretical and practical contributions. It enhances our understanding of the intersection between social media, conflict, peace, and gender issues and advances our theoretical and practical knowledge. Specifically, this study makes theoretical contributions by applying social identity theory, liberal feminist theory, and intersectionality theory in the context of social media and civil war. By exploring how social identities related to ethnicity, politics, and gender and their roles in conflict dynamics, it advances social identity theory by illustrating the role of social media in expressing and mobilizing these identities. The research further contributes to liberal feminist discourse by emphasizing the necessity of women's active participation in peacebuilding processes and highlighting the structural barriers they face in conflict-affected regions. Additionally, it enhances intersectionality theory by illustrating how intersecting identities such as gender, economic status, and ethnicity affect women's engagement with social media and their participation in peace initiatives. Broadly, this offers valuable insights into the complexities of marginalization and empowerment within conflict contexts. Integrating intersectional perspectives with the liberal feminist framework creates a more comprehensive context for analyzing the situation in Ethiopia. This expanded viewpoint can strengthen our theoretical understanding and contribute meaningfully to digital peace research.

The study offers several practical contributions to peacebuilding efforts in the digital age. First, it highlights the importance of promoting genuine dialogue and democratic engagement to transform the polarized social media landscape and foster mutual understanding. Enhancing digital literacy, particularly among women, is crucial to empower them to navigate social media safely and engage actively in discussions, thereby bridging the digital divide. Tailored digital peacebuilding initiatives can utilize the positive aspects of social media to improve communication and collaboration during conflicts. Furthermore, employing digital technologies such as artificial intelligence (AI) presents an opportunity to analyze social media interactions, identify harmful trends, support automatic moderation, and create targeted strategies for peacebuilding. Finally, integrating women's voices through inclusive peacebuilding strategies will help reduce marginalization and strengthen their contributions to stability and reconciliation efforts.

## 2 Related work

### 2.1 Social media and polarization

Understanding and addressing online polarization is crucial, as it can negatively impact mainstream politics, democratic decision-making, and society as a whole. Polarization may result in individuals encountering biased information, which can cultivate intolerance toward differing opinions, consequently leading to ideological segregation and hostility regarding major political and societal topics.[10] For this study, polarization is thus defined as animosity directed at individuals outside one's group, coupled with a sense of unity and support for those within one's own group.[11] Furthermore, social media polarization refers to the process or phenomenon in which opinions, beliefs, or behaviors become more extreme or divided, leading to a greater distance or conflict between differing groups on social media platforms.[12] It indicates the negative attitude that individuals or groups display towards individuals and groups outside their group, while also showing blind support and solidarity towards people within their group. Polarization denotes stereotyping, vilification, dehumanization, deindividuation, or intolerance of other people's views, beliefs, and identities. For this study, texts shared on social media that incite division, groupism, hatred, conflict, and intolerance are considered to contain polarization. In this study, the term polarization refers to the growing divide of opinions and political positions towards the Northern Ethiopian war disseminated on social media platforms. The term is inclusive to denote the multidimensionality of polarization such as political, religious and ethnic etc. In the Ethiopian context, group refers to ethnic, religious, political, gender or any other similar associations or identities.

Social and political polarization happens when differences between groups become very strong, leading to conflict. It makes it harder for people to connect and understand each other. This polarization is fueled by harsh language that dehumanizes others and by policies or actions that reinforce these divides. Additionally, it perpetuates the idea of perceived normative distinctions between groups, with outgroup members perceived as dangers to the survival, security, or goals of the in-group. At its worst, this kind of polarization may show up more and more as violent acts, such as assaults on opponents. The perpetuating nature of radicalization dynamics is highlighted by mis/disinformation, which both feeds into and amplifies polarization.[13]

Studies show that social media has been used as an avenue for polarization and violence.[14–16] Social media functions as a primary catalyst for politicians who, in their pursuit of power, employ disinformation to undermine their opponents by spreading misleading and manipulative content online. To this end, social media algorithms leverage sensational content to amplify false information, especially in the realm of political disinformation.[16] Through a cross-national inquiry, another study looks at how various hate speech and disinformation efforts polarize society in 177 different nations. The findings unequivocally show how hate speech and disinformation contribute to polarization.[14] Another study on US and Argentina elections documents that social media polarize voters.[17] A cross national and longitudinal study that covers 157 countries, from 2000 to 2019, examines the effects of social media on political polarization and civil conflict.[18] The findings disclose that high level of online engagement, greater social media penetration and the manners of elites use social media are related with increasing number and severity of conflicts. The study also reveals that the dissemination of disinformation correlates with increasing political polarization and which in turn increase civil conflict.

Polarization is common in virtual environments, as seen by the growing opportunities for political involvement that the digital age has brought forth. Extreme levels of politics-related rudeness have been discovered on social media platforms in different parts of the world.[19,20] A study on X conversation about the late Venezuelan president, Hugo Chávez, shows that social media users who exhibit high levels of online political polarization also tend to exhibit high levels of polarization offline.[21]

Further studies that analyze X data show that users are exposed to both people who share their opinions and those who have opposing ones.[22,23] However, exposure to opposing viewpoints does not lead to partisans becoming less committed to their positions.[24] According to this finding, X does not, at the very least, depolarize its partisan users, implying that interactions between people who hold divergent views are typically impolite and fruitless.

Therefore, the above review showcases that there is a growing interest in understanding the relationship between social media, polarization, and conflict. However, there is a notable scarcity of literature on this topic, particularly in sub-Saharan Africa. Studying the impact of social media is especially important in the African context, given the rising use of these platforms, high levels of information and digital illiteracy, as well as the infant stages of democracy in many African nations. Nonetheless, the situation in Ethiopia, remains under-researched. Studying the Ethiopian case, characterized by its diverse ethnic makeup and historical tensions, reveals broader implications for understanding complex social and political conflicts globally. The civil war in regions like Tigray highlight dynamics related to ethnic politics and governance, providing valuable lessons for conflict resolution, digital technology and peacebuilding efforts in similar contexts. As a key player in the Horn of Africa, Ethiopia's instability can have significant ripple effects on regional security, migration, and cross-border conflicts, influencing international relations and humanitarian responses. Examining the role of social media and other digital platforms in shaping narratives and mobilizing communities during these conflicts offers valuable insights for developing effective strategies to counter polarization and promote peacebuilding initiatives. These findings can inform how similar challenges can be addressed in other regions. In particular, there is a deficiency of comprehensive studies addressing the dynamics of civil war environments. Another major gap in the study of social media, polarization and conflict literature is that of the dominance of quantitative research and lack of qualitative approach. The lack of qualitative approach hinders to understand context embedded issues and perspectives and experiences of people who live in war zones about online polarization and conflict. Therefore, the study uses qualitative exploratory design to study the interface between social media and polarization in the conflict context of Ethiopia which might showcase, the sub-Saharan context. As mentioned above, this study seeks to explore the intersection of social media polarization and conflict dynamics in Ethiopia using the recent civil war in Northern Ethiopia, commonly refereed to as the Tigray War.[25],[4]

## 2.2 Women, war and peacebuilding in Africa

Many African countries continue to confront challenges, including violence and instability, yet numerous nations have made impressive strides in governance, economic growth, and social cohesion, demonstrating resilience and positive development. A comprehensive understanding of Africa requires acknowledging both the realities of conflict and the achievements made across various regions. Particularly in the Horn of Africa, numerous studies indicate that some of the most severe and protracted conflicts have taken place, highlighting the complex and enduring nature of the challenges faced in this region.[26–28] Currently, Sudan, Ethiopia and Somalia have been devastated by an ongoing conflict and civil war which shows the volatile situation of the region.

The scope, nature, and impact of the conflict's consequences vary, with human lives being permanently altered through casualties, injuries, and the displacement of individuals internally or across borders. Women and girls often bear the brunt of violent atrocities in these dire circumstances, enduring severe human rights violations and constrained opportunities due to gender disparities. Armed conflict significantly increases new infections among affected populations, with women and girls being disproportionately affected. They face increased risks of rape, and sexual exploitation, while also struggling to negotiate safe sex. These results in reproductive health complications that impact them more severely than men and boys. Gender-based violence is prevalent in these contexts, leading to profound psychosocial consequences. Additional gender-specific issues include the recruitment of girls as child soldiers and the displacement of women and girls as refugees. Access to essential public health services, particularly reproductive health care, is often inadequate in these settings.[29–31] Despite these challenges, women are crucial to the peacebuilding effort, as they represent half of the community, act as primary caregivers, serve as peace advocates, and have made significant contributions in peace processes, particularly in Sudan and Burundi, where they have participated as observers and mediators.[28,32]

The consequences of war on African women have led many to endure profoundly distressing circumstances. The work of Brittain[27] describes five ongoing effects of war that consistently impact women residing in conflict-ridden areas of the continent, including displacement, psychological and health challenges such as HIV, economic hardship, disruption of education, and sexual violence.

According to Rajivan,[33] women face systematic exclusion from the public domain, especially during times of war and in complex post-conflict settings, a phenomenon termed as "the Vicious Cycle of Exclusion". Women are often omitted from the formulation of peace agreements and reconstruction frameworks, leading to inadequate consideration of gender disparities and women's vulnerabilities in peacebuilding processes. Consequently, women's concerns are disregarded, squandering their potential contribution to peace and reconciliation efforts. The involvement of women in peace processes has been found to increase the likelihood of achieving sustainable peace. Given that women constitute a significant portion of the population, their inclusion in peace efforts is considered crucial for the success and longevity of peace initiatives.[34]

Women have a disproportionate burden of violence, human rights violations, and gender inequity. Studies suggest that while women are often targets of violent conflicts, they possess a unique capacity to mediate disputes and promote peace.[35]

Women play significant roles in the process of establishing peace. First, as peace activists and advocates, they engage in non-violent conflict resolution by fighting for democracy and human rights. Second, women help to lessen direct violence by serving as peacekeepers and humanitarian aid providers. Thirdly, women try to 'change

4 Ethiopia's devastating war.

relationships' and address the cause of violence as mediators, trauma counselors, and policymakers. Finally, women help enhance the capacity of their communities and countries to avert violent conflict by participating in education and the development process.[28] Similarly, research by Ibok and Ogar[35] underscores the significant roles women play as peace agents, showcasing their bravery and compassion in resolving conflicts where men have faltered. The study criticizes the prevalent focus in mainstream literature on portraying women solely as victims of conflict and combatants, which often overlooks the invaluable contributions women make to the peacebuilding process. Our study explores the participation of women in the peacebuilding initiatives following the devastating Northern Ethiopia War vis-à-vis its huge impact on them.

## 2.3 Digital peacebuilding

In this paper, peacebuilding entails tackling the underlying causes of conflict and promoting long-term social cohesion, development, and reconciliation in order to establish the conditions for lasting peace. It encompasses peace making and go beyond to transform the conditions that lead to conflict. Similarly, peacemaking is conceptualized as the process that employs diplomacy and negotiation to settle conflicts and bring about peace. It frequently takes place during or right after a confrontation. In the context of this study it refers the peace initiatives that enable to end the Northern Ethiopia War. However, it is good to note that peacemaking is the subset of peacebuilding, and hence sometimes we used the terms interchangeably to refer the situation in Ethiopia.

In addition, in this research, digital peacebuilding refers to the use of digital technologies, tools, platforms for resolving conflicts, fostering reconciliation, and enhancing mutual understanding among varied populations and building peace.[36,37] Thus, digital peacebuilding encompasses peacemaking and peacebuilding works using social media platforms and other digital tools such as AI technologies like natural language processing (NLP) and deep learning.[38] These sophisticated tools allow peacebuilders to efficiently collect and analyze data, address violent and divisive messages, and aid in early warning systems, conflict transformation, and transitional justice.

Technology serves a dual role in peacebuilding, both fostering connections and potentially fueling violence. Often, technology is framed in reductive terms within peacebuilding discourse, either as an inherently positive or negative force. This perspective essentializes technology by assuming its impacts are fixed and instrumentalizes it as a mere tool to serve predetermined political ends, whether for empowerment or exploitation.[39] Such a narrow view overlooks the complex ways technology interacts with social and political systems, obscuring its deeper role in reinforcing or challenging power dynamics. To address this limitation, Hirblinger et al.[36] propose shifting toward a power-conscious and reflexive analytical framework. This approach moves beyond deterministic assumptions about technology's role by critically examining how digital governance influences and is shaped by conflict resolution processes. Emphasizing considerations of power, agency, and unintended consequences, it offers a more nuanced framework for exploring the potential of technology in peacebuilding practices.

Research by Guntrum[40] illustrates how activists in Myanmar utilize information and communication technologies (ICT) to mobilize, organize, and advocate for change during crises, enabling real-time updates and strengthening community solidarity. Similarly, Sokfa[8] highlights both opportunities and risks in digital peacebuilding across Africa, such as improving communication, early warning, and peace education, while also warning of increased hate speech, misinformation, and surveillance. His study emphasizes the tension between local agency and external influence, calling for a context-sensitive, critical approach that prioritizes African perspectives. It also questions the effectiveness of relying solely on technology to resolve deep-rooted conflicts, underscoring the necessity of a nuanced and culturally aware understanding of digital peacebuilding.

The discussion above highlights that while technology improves conflict communication and coordination, its dual nature can also fuel misinformation, violence, and polarization. This necessitates strategic oversight to ensure it is used for peace rather than conflict.

## 2.4 Theoretical frameworks

### 2.4.1 Social identity theory

Developed by Henri Tajfel and John Turner in the 1970s, social identity theory offers a framework for understanding intergroup behavior and communication. It highlights the intrinsic value that individuals attach to their social group memberships and their inclination to view these groups positively. This drive for favorable group perception can result in intergroup prejudice and conflict.[41]

The process of social classification involves individuals perceiving themselves as members of specific groups, a phenomenon known as social identification. Once an individual aligns with an ingroup, they tend to seek to foster positive feelings about that group, often by evaluating their ingroup more favorably compared to other groups, referred to as outgroups.[42] As a result, the desire for positive distinctiveness for our ingroup can explain the adoption of negative beliefs and attitudes about outgroups, leading to prejudice and, ultimately, discrimination.[42]

The current research on social media polarization is connected to social identity theory through the lens of intergroup behavior and communication. Social identity theory offers insights into how individuals categorize themselves and others into social groups, illustrating how these group memberships can shape attitudes and behaviors, ultimately might be leading to polarization. In the context of social media use during wartime in Ethiopian, individuals engaging in online communities may positively align themselves with specific social groups or identities while negatively perceiving outgroups. Social media platforms often serve as arenas for individuals to express their affiliations, whether based on ethnicity, political ideology, religious beliefs, or other cultural factors. These online group memberships can become an integral part of an individual's social identity.

To show the impact of social identity on polarization, researchers West and Iyengar[43] state that behaviors observed in polarized groups such as favoritism toward one's own side and antagonism toward opposing views support the idea that social identity and group dynamics heavily influence people's political attitudes. This suggests that rather than being deliberative or based solely on objective information, political attitudes can be strongly shaped by social belonging and identity, leading to increased division and conflict among different groups.

In the context of the Northern Ethiopia War, social identity theory helps us understand how individuals and groups form identities based on their involvement in conflicting ethnic groups. During the war, media representations, especially on social media platforms, significantly affected individuals' perceptions of themselves and others based on ethnic, political and gender identities. This can also elucidate how women's roles in peacebuilding efforts are influenced by their social identities and the larger ethnic context of the conflict.

#### 2.4.2 Liberal feminist and intersectionality theories

Liberal feminism posits that women are entitled to equal opportunities in political, economic, and social domains, based on the assertion that women possess the same intellectual capabilities as men.[44] The work by Enyew and Mihrete[45] defines liberal theory as a feminist perspective that views gender inequality as a result of restricted access for women and girls to civil rights and the distribution of essential social resources, including education and job opportunities. This situation is fundamentally rooted in the socially constructed ideology of patriarchy, which maintains inequality between the two sexes. The subordination of women is often attributed to various societal and legal barriers that impede their participation and success in public life.[46] Because it emphasizes the need for equal rights and opportunities for women, liberal feminist theory is especially pertinent when analyzing women's experiences in conflict and peace processes. This viewpoint makes it easier to examine how polarization on social media can erect obstacles that prevent women from participating fully in public debate and decision-making. This theory helps explain how patriarchal norms contribute to women's marginalization and supports their empowerment by promoting greater access to opportunities and information during conflict. It also highlights the need for social and legal reforms to elevate women's voices in peacebuilding.

Liberal feminist theory emphasizes the importance of equal rights and opportunities for women, advocating for their participation in public life and decision-making processes. This framework is important for examining women's experiences during the Northern Ethiopia War and their involvement in peacebuilding efforts. Understanding how social media could empower or disenfranchise women in these efforts is made clearer through this lens, as it explores barriers women face in achieving equality within the socio-political context of Ethiopia.

The intersectionality perspective provides a vital framework for this study by revealing how different layers of identity shape women's experiences during the Northern Ethiopia War. It also helps us understand their roles and challenges in peacebuilding efforts. By examining the intersections of gender, religion, economic class, and ethnicity, historical legacies this framework elucidates how multiple forms of oppression interact to shape the lived realities of women in conflict situations.[47] This approach is particularly significant in contexts where women's multiple identities are often marginalized within broader societal discourses.

The study by Galpin,[48] highlights that social media engagement among women at the digital margins can reflect and reproduce existing power structures while simultaneously offering spaces for empowerment and agency. This duality underscores how marginalized women, especially those from diverse ethnic and socio-economic backgrounds, face unique barriers to both engagement and representation in political processes.[49] Furthermore, intersectionality informs the analysis of social media dynamics, illustrating how online platforms can both empower and disenfranchise women differently based on their intersecting identities, particularly in ways that reflect social hierarchies and inequities.[48]

# 3 Methodology

## 3.1 The research design

The study employed qualitative exploratory research design that utilizing semi-structured interviews[50] and focus group discussions.[51] Additionally, the research was supplemented by publicly available reports from mass media, NGOs and public authorities. Gender served as a key conceptual framework, directing the research focus towards understanding the nuances and impacts of gender roles within the study's context. The primary goal of exploratory research is to explore and gain insights into a problem or situations.[52,53] Thus, this study employs an exploratory approach to investigate the nexus between social media polarization, conflict dynamics, and digital peacebuilding. It specifically focuses on the participation of women in peace efforts, using the Northern Ethiopia War as a case study. Unlike quantitative research, qualitative studies provide richer insights into how information and communication technologies (ICTs) are actually applied in real-world contexts.[40]

## 3.2 Research participants and sampling

The selection of the research samples is purposive. The sample units were selected for their unique characteristics that facilitated a comprehensive investigation of the key research issues. These issues include social media polarization, conflict dynamics, digital peacebuilding, and women's involvement in peace processes. The study samples include 10 organizations that are actively engaged in peacebuilding, women's empowerment, and digital media and conflict issues (see Table 1). The research participants, including experts and directors from relevant organizations, were purposefully selected based on their experience and roles to align with our research questions. They are actively engaged with war situations, ensuring their insights are pertinent to our study. During the interviews, which took place in the participants' offices, some individuals expressed emotional responses, particularly when discussing the war's impact on women.

In the focus group discussions, we included a diverse range of participants in terms of age and gender, and facilitators ensured that everyone had an equal opportunity to contribute. This interactive environment allowed for open expression of perspectives and experiences, enriching the quality of the discussions.

Furthermore, relevant documents were collected from organizations such as the Ethiopian Human Rights Commission and International Organizations including Media. The documents are used to design interview guidelines and to substantiate the interview data. Specifically, annual and semi-annual reports from the Ethiopian Human Rights Commission, social media usage reports from the Ethiopian Media Authority, and war-related reports from international organizations were analyzed. This analysis aimed to understand the general nature of the intersection between social media, conflict, and its impact on women.

## 3.3 Data collection tools, analysis techniques and procedures

The research used semi-structured interviews, focus group discussions (FGD), and document collection as data-gathering tools. During the interview and focus group discussions, we used note-taking to capture all the required data. We use Amharic, the Federal language of the nation, and English for interviews and then the Amharic ones translated to English. We conducted ten key informant interviews and two focus group discussions (FGDs), with durations ranging from 60 to 150 min. The variation in time is due to data saturation; the length of the interviews tended to decrease towards the end as most of the research issues were adequately addressed and emerging topics became saturated. Generally, twenty research participants were involved for both categories. We organized and arranged the data for analysis, starting with transcribing interviews and typing field notes. The interview data were categorized and sorted based on the information sources, such as government offices, international NGOs, and local NGOs. Each data source, specifically an interview or focus group discussion, was assigned a unique identification number to facilitate data management, retrieval, and analysis, as can be seen in Table 1. Accordingly, the key informant interviews were coded as KII1, KII2, KII3, …, KII10, while the focus group discussions were coded as FGD, with FGD1 and FGD2 representing the two discussions.

**Table 1:** List of research participant organizations with codes.

| Organization | Coding | Types of organization | Duration |
|---|---|---|---|
| Ministry of Women and Social Affairs | KII1 | Government | 85 min |
| Ministry of Peace | KII2 | Government | 70 min |
| Ethiopian Media Authority | KII3 | Government | 105 min |
| Institute of Security Studies (ISS) | KII4 | Non-government | 100 min |
| Center for Advancement of Rights and Democracy (CARD) | KII5 | Non-government | 75 min |
| Timran Ethiopia | KII6 | Non-government | 60 min |
| Centre for Dialogue, Research, and Cooperation (CDRC) Ethiopia | KII7 | Non-government | 90 min |
| Centre for Dialogue, Research, and Cooperation (CDRC) Ethiopia | KII8 | Non-government | 45 min |
| Ethiopian National Dialogue Commission | KII9 | Government | 70 min |
| Ethiopian National Dialogue Commission | KII10 | Government | 60 min |
| The Information Network Security Administration (INSA) | FGD1 | Government | 150 min |
| Positive Peace Ethiopia | FGD2 | Non-government | 78 min |

The data was analyzed thematically. Thematic analysis is a method for identifying, analyzing, and reporting patterns (themes) within data. It minimally organizes and describes our data set in rich detail and often interprets various aspects of the research topic.[54] In this research, thematic analysis is employed to systematically identify, analyze, and report patterns such as themes related to social media polarization, conflict dynamics, and gender issues that emerge from interview and focus group data. This approach allows us to distill rich, nuanced insights into how participants' perspectives reflect broader social phenomena within the Ethiopian context of peacebuilding and conflict. We employed Williams et al.'s[55] coding procedure, which classifies coding in qualitative thematic analysis into open coding, axial coding and selective coding, to analyze and structure the data. Open coding represents the initial stage, where the researcher identifies distinct concepts and themes to facilitate categorization. During this phase, the raw data is organized into broad thematic categories to form an initial framework. Axial coding, as the second stage, builds upon open coding by refining, aligning, and grouping the identified themes more precisely. This process helps sift through and structure the data into well-defined categories, laying the groundwork for subsequent analysis. The final stage, selective coding, involves choosing and synthesizing these organized categories into cohesive and meaningful narratives, thus enabling a comprehensive understanding of the data. See Tables A.1–A.3 and Figure A.1 in Appendix A for the detailed coding, data structure, and its visualization. This approach helps in organizing and understanding the information in a way that reveals insights and deeper meanings related to the research question. The data analysis process involved repeatedly reading and comprehending all the data identifying initial themes, and subsequently determining the final themes. This is supported by the document data sources.

However, it is important to note that the aforementioned steps do not follow a strictly linear progression. Both the data collection and analysis processes were iterative. There was a continuous cycle of moving back and forth between data collection, analysis, problem re-formulation, and revising of research questions (see Figure 1). Thus, the sequence of steps mentioned does not strictly adhere to a linear structure. This approach reflects a combination of inductive and deductive processes, although the inductive process predominantly guided the work. This understanding aligns with Creswell's notion of qualitative research, emphasizing the simultaneous and iterative nature of "collecting, analyzing, and writing up the data".[52] This study also adopts the perspective of Gioia et al.,[56] viewing organizations as socially constructed entities, with members who possess awareness of their actions and intentions. This viewpoint leads us to prioritize participants' perspectives as valuable insights, rather than applying existing theories to their experiences. As a result, we focused on amplifying informants' voices during data collection and analysis to discover new concepts rather than just affirming existing ones. However, we are also mindful, as noted by

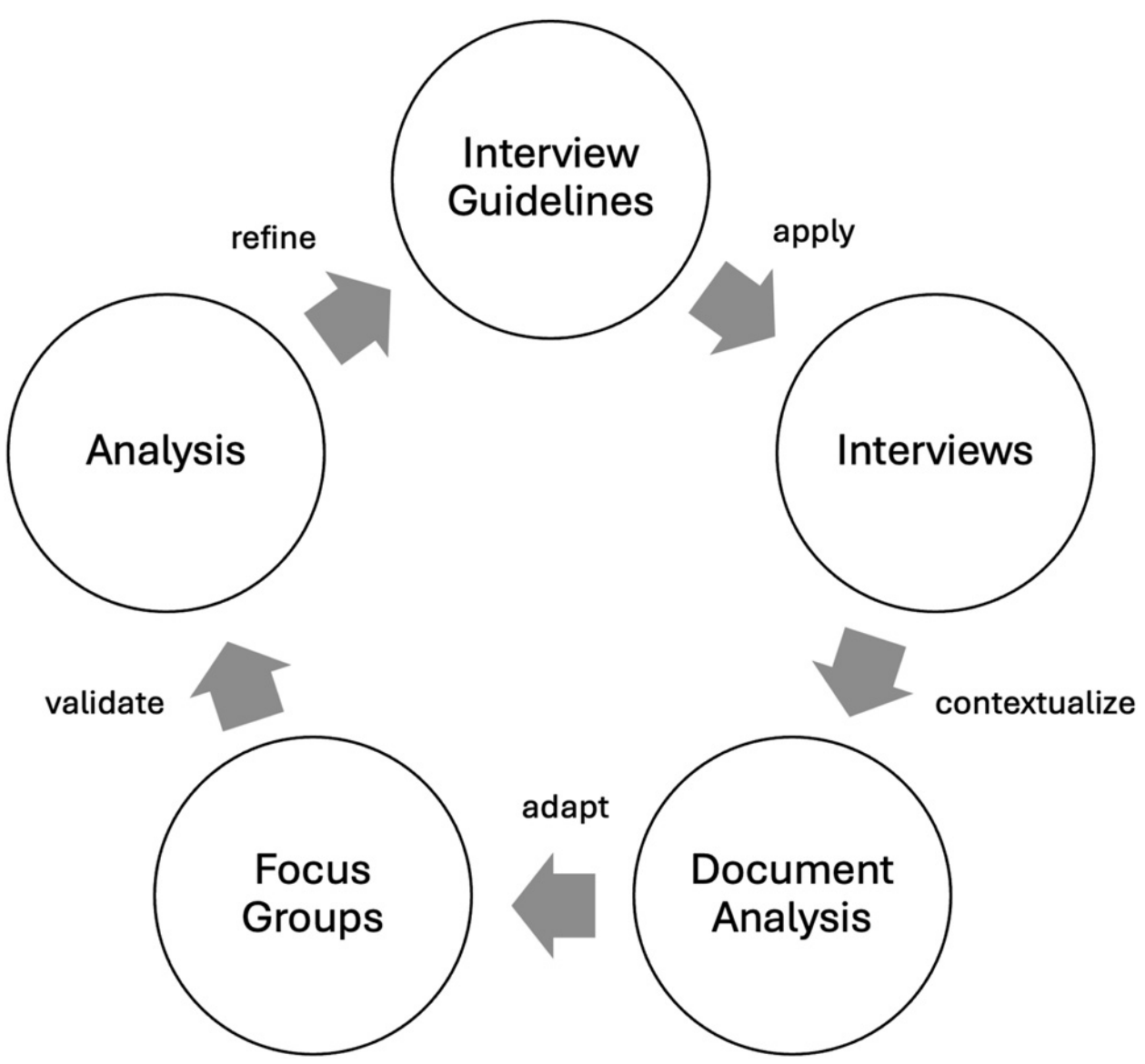

**Figure 1:** The iterative data collection and analysis process.

Gioia et al.,[56] that qualitative researchers possess a good level of knowledge and skill in identifying patterns in the data.

## 3.4 Ethical considerations

The following points were taken into account to maintain ethical standards during the data collection and reporting process:

**Permission and Informed Consent**: Prior to data collection, permission letters were obtained from Bahir Dar University in Ethiopia.[5] Continuous efforts were made to obtain informed consent from all organizations, experts, and directors involved in the data collection process. The participants were provided with a clear explanation of the study's objectives and details, ensuring their voluntary participation and understanding of the research process. Additionally, participants were assured that confidentiality would be maintained in the reporting of the study.

**Anonymity and Confidentiality**: Throughout the analysis and reporting of the findings, measures were taken to ensure anonymity and confidentiality. By maintaining anonymity, the identities of participants were protected, and their responses were reported in a way that prevented individuals from being identified. Confidentiality was also upheld by securely storing the data and ensuring that only authorized researchers had access to it. These ethical considerations were implemented to safeguard the rights and well-being of the participants, maintain the trustworthiness of the research, and adhere to ethical guidelines and principles.

# 4 Results

This section presents the findings of the study together with a detailed analysis. The data was systematically examined, leading to the identification and categorization of several key themes. Specifically, the analysis addressed: 1) the role of social media in exacerbating conflict and gender-based violence in Northern Ethiopia, 2) the persistent marginalization of women in peacebuilding and digital space, 3) the polarization and weaponization of social media within the Ethiopian context, 4) the dearth of digital peacebuilding initiatives in the country and 5) the challenges confronting digital peacebuilding efforts. Each of these themes is discussed in detail in the following sections, providing a nuanced understanding of the intersection between gender, technology, and peacebuilding in Ethiopia.

## 4.1 Social media and women during the Northern Ethiopia War

The data below reveals that the Northern Ethiopia War was exacerbated by the influence of social media and has led to a significant surge in gender-based violence against women. This violence manifests in various forms, encompassing physical, sexual, emotional, and psychological abuse. Our interviewee from the Ministry of Women and Social Affairs unveils a concerning reality that reflects the intersection of social media polarization lead war and gender-based violence amidst the war as follows:

> Sexual violence, prevalent during and after the war, remains underreported due to the politicization of data, safety concerns, and cultural taboos. Raped women face rejection and discrimination, aggravated by the absence of a digitally secured reporting system for sexual violence –KII1.

Similarly, the interview from the Ministry of Peace revealed that women were severely affected by the war in their all areas of life:

> As a human being, I have been traumatized observing the miserable situation of victims in the war regions. Words are meaningless to express the consequences of the war on women. Gang rape in front of their family members was very common, losing loved ones, including their whole family members, traumatized the victims of the war. The unbearable socio-economic crisis disturbs the lives of Ethiopian women more than their male counterparts –KII2.

The paternal system perpetuates economic dependency, severely restricting women's opportunities for financial autonomy. The FGD discussants echoed the multifaceted impacts of the war on women as follows:

> It was evident that women and children were/are affected by the war more than their male counterparts. There were reports of mass rape and gang rape, torture by inserting sticks into the genital parts, and a socio-economic crisis. And gender-based violence was the major issue that happened during the Tigray war that affected women –FGD2.

> Although we don't have an actual number, several hundreds of thousands of women were affected by the war. Women's issues were given little attention. It was underreported. During the war, civil society organizations were silenced and could not address women's issues –FGD2.

The data reveals profound trauma experienced by victims of sexual violence in conflict zones. Witnesses to these atrocities express feelings of helplessness, emphasizing the

5 We have attached a separate document.

inadequacy of language to convey the depth of suffering endured. This underscores the severe emotional distress that accompanies such violent experiences. References to gang rape and acts of torture highlight the extreme brutality of sexual violence during the war. These heinous acts inflict not only physical harm but also long-term psychological scars on victims. The occurrence of gang rape in front of family members compounds the trauma, as it is not only a victimizes the women but also deeply affects their families and communities. Furthermore, women in war-torn regions face not only immediate violence but also a worsening socio-economic crisis. This crisis disproportionately affects women compared to men, as pre-existing gender inequalities are exacerbated during conflict. Women's traditional roles as caregivers and their economic vulnerabilities contribute to their precarious situations during and after the wartime.

The acknowledgment of minimal attention given to women's issues during the war, coupled with the absence of precise statistical data, reflects a significant gap in advocacy and data collection. This suggests systemic neglect of women's experiences and needs amidst conflict, further highlighted by the silencing of civil society organizations that struggle to support victims and address gender-based violence. Women and children suffer disproportionately in the war, facing gendered violence that targets them both as individuals and as symbols of community, heightening their vulnerability. While exact numbers remain elusive, claims that "hundreds of thousands of women were affected" indicate a widespread crisis that transcends individual experiences. The lack of quantitative data contributes to the ongoing underreporting of sexual violence and could complicate the ability of policymakers and aid organizations to respond effectively.

Overall, the integrated data paints a troubling picture of the intersection between war and gender-based violence, emphasizing the trauma suffered by women and children in conflict zones. Systemic issues, including socio-economic hardship, underreporting, and insufficient attention from civil society, further compound the suffering of victims. This analysis highlights the urgent need for robustness reporting systems, targeted interventions, and a greater focus on women's issues in humanitarian responses to conflict. Addressing the systemic neglect and silence surrounding these issues requires a concerted effort from various stakeholders, including governments, NGOs, and international organizations, to advocate for the rights and needs of women affected by war. This reporting gap reveals the lack of a secure digital reporting system specifically designed to address women's issues.

Moreover, the data below revealed social media, particularly Facebook, X, and YouTube, were used to aggravate the war in general and gender-based violence in particular by disseminating polarized messages. The following interviewee expresses the extent of social media use as a weapon of war:

> Social media was used for fuelling conflict. Especially Facebook, X, and YouTube were widely used during the Tigray war. I suspended my Facebook account during the war to get mental peace, since there were several horrific video posts widely circulated in the country that aggravated the war –KII7.

The language of social media was very volatile and divisive as we can understand from the following data:

> Social media users were used divisive terms such as junta, genocide, conqueror, devil, day heyna, killer, fascist, slaughterer to refer to warning groups, ethnic groups and political parties. For example, it was very common to refer to the TPLF as junta and day hyena, that latter extends to refer to the Tigrians –KII2.

Our data revealed that social media was used as weapon. The weaponization of social media (see Section 4.3) has exacerbated gender-based violence, including the spread of ethnic hate speech, threats, and targeted attacks against ethnic groups. Social media was used as a propaganda tool and media reports of rape were unethical, which affected the dignity and security of women.

The data presented above reveal the extensive consequences of the social media-fueled conflict on the lives of women. These impacts have led to significant and enduring challenges for women, encompassing a range of hardships. Similarly, the annual report of EHRC[6] highlights the widespread and organized gender-based and sexual violence inflicted on women and children during the conflict in Northern Ethiopia. The study highlights how social media has been instrumental in exacerbating the conflict in Northern Ethiopia, and women were disproportionately impacted by the conflict, enduring a wide array of physical, sexual, psychological, and socio-economic repercussions. In this context, women and children are recognized as the most vulnerable groups in African societies, often inadequately prepared for, affected by, and affected during civil wars, violent conflicts, genocides, and upheavals.[57]

The data on the use of social media during the Tigray conflict highlights its significant role in escalating tensions and raises ethical concerns about the representation of vulnerable populations. Platforms such as Facebook, X (formerly Twitter), and YouTube facilitated the rapid dissemination of information, often intensifying conflicts

**6** https://shorturl.at/PgcdU.

instead of fostering understanding. Users reported suspending their Facebook accounts for "mental peace," reflecting the psychological toll of exposure to graphic content and negative interactions. Inflammatory language used to describe ethnic groups and political factions further emphasizes a culture of hostility, with derogatory terms like "junta" and "genocide" contributing to the dehumanization of opponents and potentially inciting violence.

Additionally, the media's role as a propaganda tool illustrates how news coverage can be manipulated for political agendas, distorting public perception and polarizing communities. Ethical violations in reporting sexual violence highlight serious concerns regarding women's dignity and security (FGD2). Such reporting often fails to respect the dignity of victims, leading to re-traumatization and further stigmatization. This underscores the need for responsible journalism, particularly in conflict scenarios involving sensitive topics like sexual violence. While both social media and traditional media can inform and connect communities, they also risk amplifying conflict and undermining the dignity of marginalized individuals. This analysis emphasizes the urgent need for ethical media practices and greater awareness of the impact of language and imagery to foster constructive dialogue and promote healing in post-conflict settings.

## 4.2 The marginalization of women in peacebuilding and digital space

In order to understand the inclusivity of the Ethiopian peacebuilding process, we asked to know the extent of women's involvement in the peacemaking that enable to end the 2 years war vis-à-vis the heavy burden of the war on them. In the negotiation and signing of peace initiatives that aimed to halt the violent conflict in Northern Ethiopia, women were not involved, which highlights the exclusion of their perspectives. The following data indicates that women were marginalized in both peacemaking and the digital media space due to various factors, including sociocultural norms, household responsibilities, economic challenges, media illiteracy, and the digital divide:

> Efforts to involve women in peacebuilding processes, including UN Women-supported conferences, have unfortunately not yielded observable impacts; despite the higher expectation of success. Furthermore, women were excluded from the peace agreements. No single woman participated in the peace initiatives signed in Pretoria and then Nairobi –KII1.

> In the Pretoria Peace Agreement, the absence of female representation raises concerns, impacting the role of women in the national dialogue that aspires to build sustainable peace in the country. Security reasons and cultural barriers hinder the active participation of women in such type of critical issues in building peace –KII9.

The above data show that women were marginalized in the peacemaking that end the war. This marginalization might affect the participation of women in peacebuilding activities. This discloses the gender bias of the peace agreements and the peacebuilding initiatives are marginalizing women's meaningful participation from the initial stages of peacemaking. The peacemaking initiatives were top-down. The top-down peacebuilding initiatives are exclusively dominated by males. In their absence, it's hard to get women voice heard and to include their perspectives. The top-down approach excludes the historically marginalized women from directly participating in the peace process. Despite women being represented in the government, crucial decision-making still is done by male officials. This masculine-centered peacemaking approach marginalized women and obliged them to be voiceless while deeply affected by the war. This marginalization mirrors the traditional gender roles in the country that consider war and peacemaking issues to be the duty of males. In Ethiopia, women have historically occupied subordinate positions characterized by male dominance, facing cultural discrimination and experiencing limitations in their participation in warfare and peacebuilding processes.[58] The exclusion of women from peace processes can be understood as a social norm rooted in patriarchal structures, which limit their opportunities for meaningful involvement in peace and reconciliation initiatives.

Restricting the involvement of women in peacemaking efforts means the exclusion of their perspectives. This is preventing them from shaping and tackling their issues and disregarding their insights and understanding of the conflict at hand from public discussions. Marginalizing more than half of the total population will halt sustainable peacebuilding in the country.[7] The potential of women to mediate conflicts and facilitate peace is often underestimated. Recognizing the critical role women can play in peacebuilding processes challenges the view that excludes them and highlights their importance as key stakeholders in conflict resolution.[28]

The exclusion of women from peacemaking efforts in the context of social media polarization and weaponization in Ethiopia can be attributed to various factors. When we

7 https://population.un.org/wpp/.

asked why women were marginalized in the peacemaking, Our interviewees answered as follows:

> I believe the patriarchal societal system is the root cause of the marginalization of women. Because of this, women were excluded from the peace deal –KII7.

> Women's participation in digital platforms is hindered by household burdens, and media illiteracy, limiting their engagement with the digital world. This participation in the digital media discourses is very limited –KII10.

We can summarize the major factors that hindered women's participation in the social media environment as follows:

**The hostile online environments**: The hostile online environments discourage women from actively participating in digital peace-building initiatives, as they feel unsafe and unprotected. for example, our data disclose that rape was encouraged as a revenge strategy by war combatants.

**Digital divide**: Women in Ethiopia, particularly those in rural areas and marginalized communities, often face barriers to accessing technology and lack adequate digital literacy skills. This digital divide further marginalizes women, making it difficult for them to engage in online discussions and participate in peace-building efforts. It is safe to argue that the social media ecology is dominated by males in Ethiopia.

**Cultural and societal norms**: Traditional gender roles and societal expectations can hinder women's participation in public discourse, including peace-building initiatives. Deeply ingrained cultural norms may discourage women from speaking up or taking leadership roles in these contexts, even on digital platforms. The cultural and societal norms in Ethiopia favor males speaking publicly and encourage women to be shy in the public arena. These traditional societal norms discourage the engagement of women in the digital world and make them remain marginalized in peacemaking activities.

The data underscores significant concerns regarding the exclusion of women from peacebuilding processes, illustrating the disappointing outcomes of efforts to involve them in these critical initiatives. Despite the support from organizations such as UN Women, conferences aimed at integrating women into peace efforts have not produced observable impacts or met the expected success. This gap is particularly evident in high-profile peace agreements, such as those reached in Pretoria and Nairobi, where not a single woman was represented in the negotiations. This lack of female participation not only raises questions about the inclusivity of these agreements but also diminishes the potential for sustainable peace, as women's perspectives and experiences are essential in national dialogue and conflict resolution. Furthermore, the absence of women in the Pretoria Peace Agreement poses substantial concerns for future peacebuilding efforts. Their exclusion from decision-making processes undermines attempts to create a truly representative and effective approach to peace. The patriarchal societal system is identified as a root cause of this marginalization, suggesting entrenched gender biases that preclude women from contributing to peace deals. Cultural barriers and security concerns are significant factors limiting women's active participation in such critical discussions, exacerbated by household burdens and media illiteracy that hinder their engagement with digital platforms. Consequently, women's participation in digital media discourses remains very limited, further isolating their voices from essential discussions.

This systemic exclusion highlights a broader issue regarding societal norms and structures that continue to marginalize women in conflict resolution. Ultimately, the absence of women from peacemaking initiatives not only underscores the urgent need for inclusive strategies but also emphasizes the necessity of addressing underlying patriarchal systems that inhibit women's active involvement. Recognizing and harnessing the vital contributions women can make is essential for achieving lasting peace and fostering an inclusive environment that values diverse perspectives in the peacebuilding process.

## 4.3 Social media polarization and weaponization

Social media polarization is the phenomenon where individuals and groups on social media platforms become increasingly divided into opposing camps with strongly held views, often leading to heightened conflict, hostility, and a lack of constructive dialogue.[59] The data below indicates that social media served as a dual-purpose tool for both polarization and weaponization throughout the conflict in Northern Ethiopia:

> Social media platforms, particularly Facebook, played a detrimental role in disseminating misinformation and hate speech, intensifying the war. Posts propagating violence highlight the influence of social media on polarization and weaponization. Divisive language use such as we and they to refer to warring and ethnic groups and political parties was very common among social media users. Some social media users were calling to kill and arrest all Tigrians by blaming them as junta and opportunists. And some Tigray activists were calling the killing of Amhara by blaming them as conquerors of their land and supporters of the Tigray genociders –KII1.

> Hate speech and polarization were characteristics of the social media environment in the country. For example, it was very common to call TPLF as junta, and day-hyena which was equally used to refer to the Tigiray people regardless of their political

> affiliation by the government affiliated social media users. In the man time, it was also common to call the Ethiopian government as fascist and the Ethiopian army and the allied forces as genociders by the TPLF affiliated social media armies. –KII5.

> The media, particularly social media, is identified as contributing to warlike sentiments and tragic incidents during the war. Digital peacebuilding efforts are notably absent, emphasizing the urgent need for a well-designed strategy and implementation tools in the digital realm –KII10.

> The polarized media environment distorted the reality of the war. Social media was used to aggravate conflict, not peacebuilding. Social media was used to fuel conflict –FGD2.

> The presence of low media literacy skills is identified as a significant contributor to social media polarization, hate speech, and disinformation. Regrettably, digital peacebuilding efforts are currently non-existent –KII1.

The data reveals the significant and detrimental role of social media platforms, particularly Facebook, in fostering division and disseminating misinformation during the Tigray conflict. Misinformation and hate speech proliferated, intensifying polarization and violence among communities. Users frequently employed divisive language, referring to warring ethnic groups and political parties as "we" and "them," which further entrenched animosities. Notably, some social media users initiated violence against Tigrians, labeling them as "junta" and "opportunists," while Tigray activists retaliated with similar calls for violence against Amhara individuals, denouncing them as conquerors and supporters of genocide. This cycle of hate amplified a polarized environment where derogatory terms were common; for instance, the TPLF was frequently termed "junta" and "day-hyena," illustrating the dehumanizing rhetoric prevalent across social media platforms. Simultaneously, TPLF-affiliated users characterized the Ethiopian government and military as "fascist" and "genociders," reflecting how deeply entrenched and contentious narratives contributed to an overall environment of hostility. The prevalence of misinformation emerged as a primary driver of offline conflicts, surpassing even hate speech in its impact before, during, and after the war. The media landscape, characterized by sectarianism and polarization, failed to provide constructive dialogue, with mass media often mirroring the divisive sentiments prevalent on the offline environment.

The research findings underscored that social media represents a virtual stage of warfare, employed as a mechanism for spreading misinformation and disinformation throughout the conflict. Similar studies such as those by Haile,[60] Wassie et al.[61] enhanced the findings about social media polarization and its use for fuelling conflict (weaponization) in Ethiopia.

## 4.4 The dearth of digital peacebuilding in Ethiopia

This section delves into the dearth of digital peacebuilding in utilizing digital tools to promote peace and address conflicts in Ethiopia. Our study revealed the dearth of digital peacebuilding in Ethiopia vis-à-vis the widespread digital media polarization and weaponization.

Key informants (KII10) emphasize that social media is a critical factor in exacerbating tensions, with statements indicating that the platforms are primarily serving as catalysts for conflict rather than promoting reconciliation or peace. KII10 notes that while social media is instrumental in fueling conflicts, there are no significant digital peacebuilding efforts underway to counteract this trend. This sentiment is echoed by multiple informants, including KII3, KII9, KII10, and KII2, who collectively assert that "the role of the digital media in the peacebuilding process is almost zero," with phrases illustrating the severity of the issue.

Furthermore, there is an alarming observation that a small number of digitally connected individuals disproportionately influence the broader, digitally non-connected population, further disturbing national peace and stability. KII9 remarked that "the few digitally connected persons affected the digitally non-connected wider population and disturbed the peace of the nation." Statements such as "the media, especially social media, are warmongers," from KII3, and "social media is the major problem of the nation," from KII2, emphasize the critical view that the prevailing use of digital platforms contributes significantly to societal unrest. This portrayal of social media as aharmful force underscores the potential for misinformation, hate speech, and divisive rhetoricescalate conflicts rather than foster dialogue and understanding.

The data reflects a pressing need for the development of digital peacebuilding strategies that can harness the potential of social media to promote peace and reconciliation, rather than allowing it to remain a tool for discord. The findings call for a concerted effort to establish frameworks that encourage positive engagement through digital media and work toward healing the divisions within society.

There are active developments in digital peacebuilding that harness the transformative potential of technology to address conflicts and promote peace. The tools and strategies are continuously evolving, and some successful implementations can offer inspiration for Ethiopia. A notable example is Ushahidi,[8] an open-source platform widely used in Kenya. It enables citizens to crowdsource data, report incidents, and map events related to violence and peacebuilding efforts, leveraging accessible technologies such as mobile phones and the Internet. Tools like Ushahidi amplify voices, empower individuals, and facilitate community mobilization through versatile data management and analysis capabilities.

Drawing upon the principles highlighted in the INEF Report by,[38] integrating AI technologies – such as social natural language processing (NLP) and deep learning – can significantly enhance digital peacebuilding efforts. These advanced tools enable peacebuilders to quickly gather and analyze data, respond to violent and polarized messages, and support early warning systems, conflict transformation, and transitional justice.

For Ethiopia, adopting such models and utilizing AI-driven solutions could greatly advance digital peacebuilding initiatives. This approach would address social media polarization, empower women, and create peace initiatives tailored to the nation's unique dynamics. By learning from successful implementations like Ushahidi, Ethiopia can move towards a more structured and effective digital peace landscape, supporting reconciliation and enhancing societal harmony.

## 4.5 The challenges of digital peacebuilding in Ethiopia

As mentioned above, digital peacebuilding in Ethiopia is at the infant stage. It encounters numerous obstacles, much like in other sub-Saharan countries. Some specific challenges faced in Ethiopia include (see Figure 2):

1. Digital Divide: There is a significant disparity in internet access and digital literacy across Ethiopia. The divide between men and women, urban and rural areas, along with differences related to income and education, can restrict the effectiveness and outreach of digital peacebuilding initiatives. In Ethiopia, the limited number of digitally connected people affects millions of lives for those who are not connected, which shows the interface between the online and offline socio-political environment of the country. As our interviewee states (KII8) "The digital conflict, while having limited coverage, manifests widespread offline consequences, particularly affecting digitally illiterate masses across the country". In addition, one can see the digital gaps in the country from the limited number of social media users, as Digital Ethiopia shows that only 5.5 % of the population uses social media.[9]

   However, our data shows social media is one of the factors that cuts the social fabric of the community and fuels conflict during the Northern Ethiopian war and beyond which affects the majority of the population in Tigray, Amhara, Afar regions as well as conflict in Oromia region. As KII1 mentions, "Social media create division, sow suspicion, destabilize the harmonious social relations among the different ethnic groups in Ethiopia, and fuel conflicts".
2. Government Control and Shutdown: The Ethiopian government has been known to limit internet access and block online content, and shutdown internet, which can stifle free expression and undermine the impact of digital peacebuilding efforts. Particularly, the repeated shutdown of internet service in regions such as Tigray, Amhara and Oromia for long period of time (KII2 and KII5) potentially halts digital peace building initiatives.
3. Ethnic and Political Tensions: Ethiopia's complex ethnic and political environment, alongside its historical conflicts, often carries into the digital space (KII1, KII2 and KII4). Social media is dominantly utilized to spread hate speech and false information, worsening conflicts instead of fostering peace. This is consistent with other studies, such as,[62,63] that underscore the dissemination of ethnic hate speech in Ethiopian social media, landscape.
4. Security Concerns: Threats to cybersecurity, including hacking, online monitoring, and data breaches, present risks to digital peacebuilding efforts, compromising the safety and privacy of those involved. For example, our data discloses the absence of a safe digital reporting systems for women who face sexual violence (KII1).
5. Lack of Trust and mis/dis Information Circulation: Establishing trust in online platforms for peacebuilding is difficult, especially in an atmosphere rife with dis/misinformation. Ensuring that the information shared online is accurate and trustworthy is vital for effective peacebuilding. In Ethiopia, during the wartime, information warfare was used as a war strategy by both parts, in which dis/misinformation was widely circulated in social media platforms such has Facebook, X and YouTube (KII1, KII2, KII3).
6. Capacity Limitation: Enhancing the digital skills and capacities of those involved in peacebuilding is crucial. Absence of adequate training programs and resources

8 https://www.ushahidi.com/.

9 Digital Ethiopia-2024: https://datareportal.com/reports/digital-2024-ethiopia.

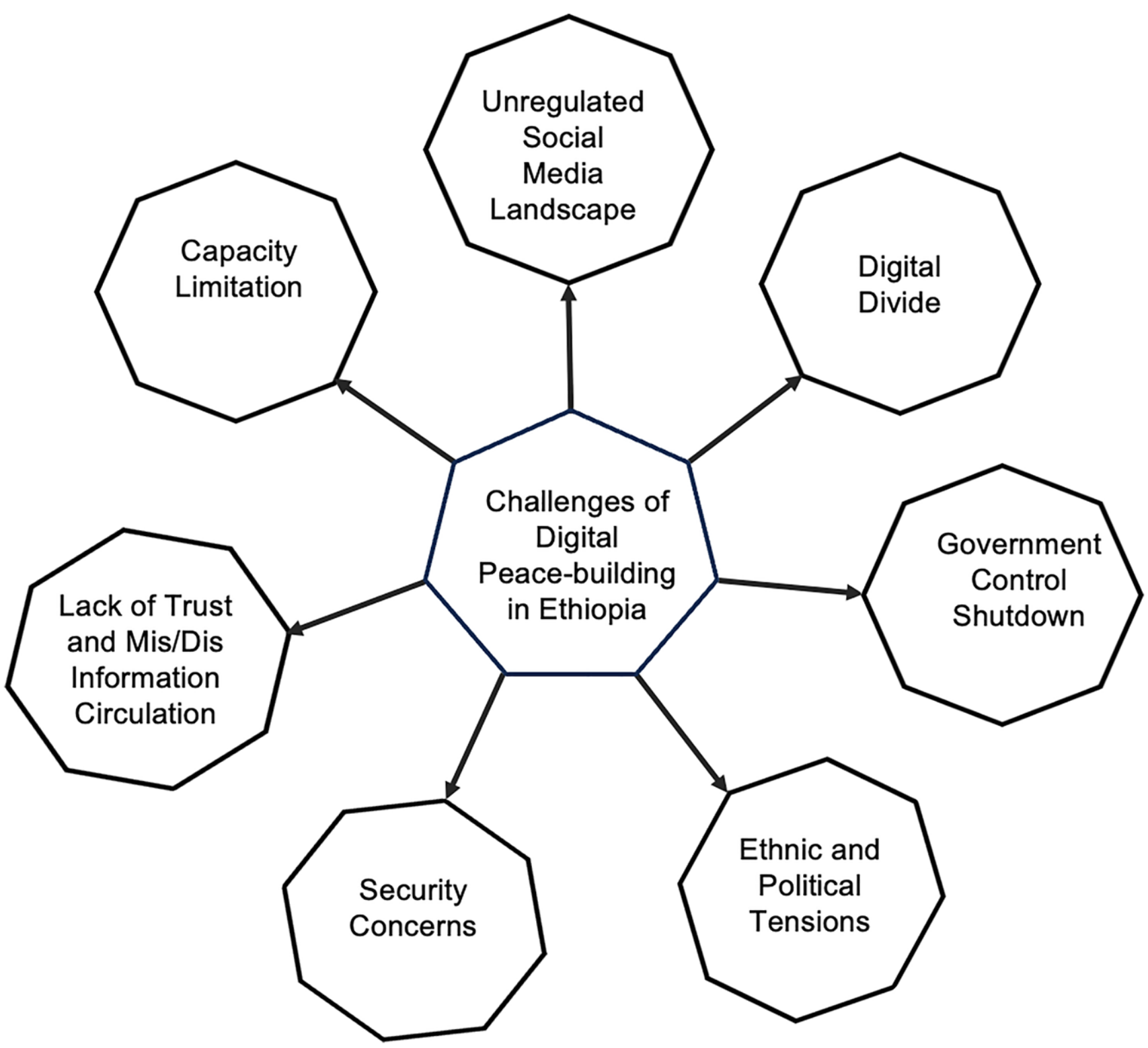

**Figure 2:** The challenges of digital peacebuilding in Ethiopia.

hinders stakeholders to effectively utilize digital tools for peacebuilding (KII1 and FGD1). Specially, offices such as the Ministry of Women and Social Affairs and Ministry of Peace lacks the required resources and skills to use the digital media for empowering women's and peace building. For example, utilizing Natural Language Processing (NLP) and machine learning tools for thorough data analysis have the potential to mitigate the negative impacts of social media within the nation. Tools for data visualization and categorization, could serve as an effective intervention to counteract the adverse effects of social media. Our research underscores the deficiency of such technological tools in the underutilized and inadequately understood realm of digital peacebuilding in the country.

7. Unregulated Social Media Landscape: The social media platforms are not adequately regulated by the major social media giants such as Meta and X (FGD1 and KII2), which creates favorable condition for polarization and weaponization. The algorithm design enhances echo chambers that reinforce polarization and weaponization of social media, especially during the war context. Besides, the findings elucidate that the Ethiopian hate speech and disinformation regulation couldn't enable individual account users to be accountable for their hatful and weaponized messages. The lack of instructional manuals to guide its implementation impedes the proper enforcement of the regulation. There is also a confusion and policy gap about who is responsible for implementing the regulation, and hence so far there is no significant effect in normalizing the social media environment since its ratification in 2020 (KII3).

## 5 Discussion

This discussion interprets the findings of the study concerning the intricate interplay between social media, polarization, and conflict in the context of Northern Ethiopia. It highlights how digital platforms serve to both reinforce ethnic and political identities and escalate violence through processes such as vilification, dehumanization, and the weaponization of narratives. The analysis is framed by three key theoretical perspectives: social identity theory, which explains how group categorization fuels hostility; liberal feminist theory, which underscores the exclusion and marginalization of women in peace and conflict processes; and intersectionality theory, which emphasizes how overlapping social identities amplify vulnerabilities and influence conflict dynamics. Together, these frameworks provide a comprehensive understanding of how social media influences conflict escalation, mobilizes ethnic politics, and marginalizes women's participation in peacebuilding efforts. The discussion also considers the implications for digital peacebuilding strategies and the critical need for inclusive, gender-sensitive approaches in resource-constrained, conflict-affected settings.

In today's digital age, social media has emerged as a powerful tool that can either promote dialogue and understanding or exacerbate divisions and fuel conflict. In the context of the recent Northern Ethiopia conflict, social media has played a significant role in shaping and intensifying divisions among various ethnic and political groups. The empirical data collected through interviews and focus group discussions reveal patterns of polarization consistent with social identity theory.[41] Our findings reveale that social media platforms such as Facebook, YouTube, X, and Telegram serve as arenas where factions amplify in-group loyalty and out-group hostility through tactics such as vilification and dehumanization. These behaviors align with social identity theory assertion that categorization into in-groups and out-groups reinforces group loyalty while dehumanizing adversaries, thus facilitating conflict.

Through vilification, dehumanization, extreme language use, and invalidation, social media users have escalated tensions and deepened hostilities. This behavior exemplifies the desire for positive distinct identity as individuals seek to enhance their social identities by negatively portraying rival groups. Vilification serves as a strategy through which users involved in the conflict demonize their adversaries, reflecting the insights of McKinley, who notes that the classification of individuals into groups leads to social identification and prejudice.[42] Rival groups use exaggerated and inflammatory language to cast their opponents in a negative light. Terms such as "junta," "killer," "genocide," "mass gang rape," and "human slaughter" are frequently employed to exaggerate the actions of rival factions, thereby inciting emotions among supporters. Such vocabulary not only stokes fear but also serves to justify extreme reactions against the opposing group by framing them as an existential threat. The terms of reference have been echoed by the interviewees and FGD discussants, who noted that such vocabulary is commonly used on platforms like Facebook and X, YouTube. This vocabulary serves to mobilize collective support and reinforce in-group identity while casting rival groups as existential threats, consistent with the mechanisms described by social identity theory. These expressions serve not only to mobilize collective support but also to reinforce ingroup identity by framing rivals as existential threats. This aligns with the assumption of social identity theory that categorization into groups fosters in-group loyalty and facilitates out-group hostility.[41] Such language contributes to a cycle where negative portrayals of opposing groups diminish empathy and justify extreme measures, including violence.[42] By explicitly linking the observed language use to social identity theory, this analysis demonstrates how online narratives sustain polarization and conflict escalation.

Furthermore, the practice of dehumanization has become prevalent, as users resort to derogatory terms to weaken the perception of humanity their foes. This practice reduces the perceived moral boundaries between 'us' and 'them', thus enabling extreme reactions and violence. As social identity theory suggests, reducing out-group empathy facilitates the escalation of hostility and conflict.[41] Labels such as "day-hyena" and "devil" were widely used to strip individuals of their personality and morality, making it easier for members of one group to rationalize violence against another.

The use of extreme language during the conflict further illustrates the volatile environment fostered by social media. Statements like "we should not allow TPLF to live again" and "they should be demolished from existence" exemplify the toxic rhetoric that spreads in online discussions. Such extreme expressions reflect deep-seated animosity and confirm the role of social identity in perpetuating polarization, legitimizing calls for violence and the eradication of entire groups. The interviewee data reveal that the "us versus them" dichotomy is a direct reflection of social identity dynamics, which has been a pervasive theme across social media platforms. This binary perspective effectively distances and segregates the various ethnic and political groups involved in the armed struggle, focusing on Tigray, Amhara, Afar, and Oromo ethnic groups, as well as the political parties established along those lines, such as the Tigray People's Liberation Front (TPLF), Amhara National Movement (ANM), and Oromo Liberation Front (OLF). This framework encourages the demonization of the other side while fostering a strong sense of ethnic and political loyalty among supporters. From a theoretical perspective, this dynamic illustrates how social identity and categorization reinforce conflict and polarization, often leading to dehumanization and violence, as predicted by social identity theory.[42] Those who adopt this perspective are conditioned to view any interaction with members of the opposing side as a betrayal, reinforcing the divisions cultivated by social media.

Furthermore, calls to isolate the Tigray regional state from Ethiopia have been prevalent on social media platforms, with supporters arguing that the Tigray people should not coexist with those they label as "genociders." Such rhetoric not only deepens societal polarization but also hampers peace and reconciliation efforts, demonstrating how social identity and group dynamics can escalate conflict rather than foster understanding. This situation in Ethiopia illustrates how social media can be exploited to fuel division and violence. In contrast, Gichuhi[64] observes that in West Africa, social media predominantly functions as a tool for peacebuilding, serving various purposes such as raising awareness, conducting campaigns, sharing narratives, providing training, and combating hate speech online. The comparison emphasizes that the role of social media is highly context-dependent, influenced by political and social factors within each region, which shape whether these platforms contribute to conflict escalation or peace promotion.

The role of social media in disseminating and amplifying polarizing narratives in Ethiopian context reinforces the cyclical nature of conflict between the offline and online environments, one reinforces the other. This is consistent with other studies that disclose the interface between social media polarization and conflict.[14–16] It is good to argue that the polarization of social media in Ethiopia can be attributed to the political economy of mass media in the nation, mirroring the broader political economy of the state. The broader political economy of the state is shaped by ethnic politics and divisive discourses. This argument aligns with findings from studies by Dessie et al.,[65] Skjerdal and Moges[66] that state ethnic polarization and ethnicization of mass media in Ethiopia affect the media landscape of the country. Social Identity Theory is crucial for understanding social media polarization in Ethiopia, where ethnic identity plays a significant role in shaping the nation's politics, society, and economy. In Ethiopia, individuals often categorize themselves based on their ethnic affiliations, which promotes a strong ingroup identification and can lead to bias against outgroups. This ingroup favoritism manifests in social media interactions, where users may amplify narratives that benefit their ethnic group while trivializing others, contributing to an increasingly polarized environment. As people engage within echo chambers that reinforce their existing beliefs, the resulting hostility can escalate tensions between ethnic groups, ultimately impacting national cohesion.[67] Understanding these dynamics through the lens of Social Identity Theory is essential for addressing the issues of prejudice, discrimination, and conflict that arise in the intersection of social media and ethnic identity in Ethiopia.[42]

As illustrated in Figure 3, social media polarization has led to its weaponization, fueling conflict during the War and contributing to gender-based violence, including killing, displacement, and sexual harassment of women in war zones. Another study[68] also reveals multiple forms of violence including sexual, physical, and psychological violence, with reports of gang rapes of minors as young as 14 and sexual violations of pregnant women and elderly women up to 65 years old seeking refuge in religious institutions.

It is sound to argue that this situation is aggravated by the inadequate digital peacebuilding, which leaves social media polarization and weaponization unaddressed. Digital peacebuilding employs advanced technologies such as natural language processing (NLP), large language model (LLM), topic modeling, and hate speech detection to mitigate these effects. These tools can effectively identify and counter harmful narratives, fostering constructive engagement and promoting inclusive dialogue. In this regard Sokfa[8] states that the growing use of open source technology in digital peacebuilding in Africa is driven by its adaptability, accessibility, and cost-effectiveness. From a theoretical perspective, integrating these digital tools into peacebuilding efforts can be seen as a practical application of conflict transformation approaches, helping to address the root causes of polarization.

The findings further reveal that women were marginalized in the peacemaking and peacebuilding process. Restricting women's participation in these efforts results in the exclusion of their perspectives. This illustrates the "vicious cycle of exclusion" defined by Rajivan,[33] where women are sidelined from developing peace agreements and reconstruction plans. As a result, insufficient attention is given to gender disparities and the specific vulnerabilities women face in peacebuilding endeavors. Incorporating feminis theories into this analysis emphasizes the necessity of gender-inclusive approaches for sustainable peace, aligning with the broader goals of social justice and equality.[69]

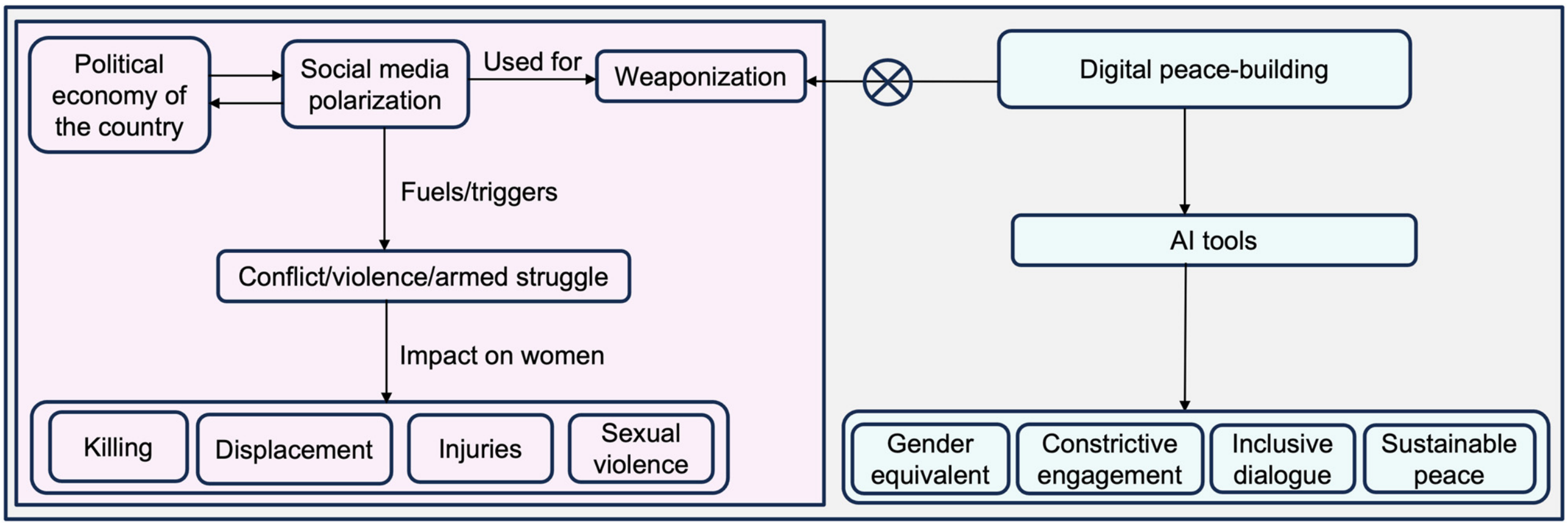

**Figure 3:** Social media polarization and the dearth of digital peacebuilding.

The marginalization hinders their ability to address and resolve their own challenges and disregards their insights into ongoing conflicts in public discourse. Liberal feminist theory asserts that patriarchal structures interfere with women's personal and political decisions.[69] Marginalizing more than half of the population will impede sustainable peacebuilding in the country. This could halt the country's performance towards achieving sustainable development goal (SDG) 5, which focuses on gender equality, and SDG 16, which pertains to peaceful and inclusive societies for sustainable development.[70] Overlooking their unique ability to mediate conflicts and promote peace, as highlighted by Ibok and Ogar,[35] Lisa and Manjrika[28] who emphasize women's capacity to foster peace through "courage and love," the crucial role of women in peace processes is ignored. This demonstrates the importance of integrating gender perspectives into peace and conflict theories to better understand and address the specific needs and contributions of women in conflict settings.

The combined impact of the online and offline conflict convergence negatively affects the nation as a whole, with a particular emphasis on women. The intersectionality perspective reveals that various identities, including gender, socio-economic status, and ethnicity, amplify the challenges women face during peacemaking efforts.[47] This highlights that the visible digital conflict occurs amidst a lack of organized digital peacebuilding initiatives. An alarming need for digital peacebuilding tools is evident, especially considering that the majority of government offices such as the Ministry of Women and Social Affairs, the Ministry of Peace, the Ethiopian National Dialogue Commission, etc. lack capacities to use the digital space for nurturing peace and harmony. Applying intersectional feminist theory elucidates how overlapping social identities impact women's experiences of conflict and peacebuilding, emphasizing the need for contextually sensitive interventions.[47]

Digital technologies are pivotal in peace-building and women's empowerment by integrating women's perspectives into all stages of digital peacebuilding. Ethiopia has yet to fully leverage advancements like LLMs and machine learning to address social media polarization and prevent its weaponization. AI can facilitate trend prediction and identify polarized content, using NLP techniques such as topic clustering and named entity recognition to mitigate adverse social media impacts. Its clear to argue that utilizing AI-based conflict analysis models aligns with conflict transformation and peace building theories that seek root cause resolution through data-driven insights. Machine learning enables in-depth data analysis for informed interventions. Tools like Ushahidi empower communities by turning citizen-generated data into actionable solutions through intuitive crowdsourcing and mapping capabilities.[8]

Our research highlights the significant gap in the availability and application of such technological tools, which hampers Ethiopia's digital peace efforts. Moreover, digital peacebuilding efforts are challenged by high levels of digital divide, government control and shutdown, ethnic and political tension issues, lack of trust, capacity limitation, and unregulated social media landscape. addnewWhile these constraints limit AI deployment, understanding these structural barriers from a theoretical perspective, such as digital inequality and access, shows that technological solutions alone are not enough. They need to be complemented by broader socio-political reforms to effectively support peacebuilding efforts. From a sociotechnical perspective, digital systems are embedded within complex networks involving both human and non-human elements, such as algorithms, platforms, and artifacts. These interconnected components influence social interactions, discourse, and power dynamics through their design and use.[71,72] This is similar with Soka[8] research that reveals structural barriers, including unequal internet access, low digital literacy, and government-imposed restrictions like internet shutdowns, further limit the effectiveness of digital peacebuilding efforts.

The findings further highlight that the impact of a minority of negative online actors on the lives of millions of Ethiopians without internet access, particularly affecting women, remains a pressing concern. The potential harm extends beyond digital realms, emphasizing the need for comprehensive strategies that account for the broader societal implications of online activities. Social media algorithms tend to prioritize sensational content, thereby amplifying misinformation, especially within the context of political disinformation.[16] This phenomenon highlights the importance of algorithmic accountability and the role of ethical AI development in conflict and peace studies. Other studies also enhance the intersection between the online and offline environment.[21,24,73]

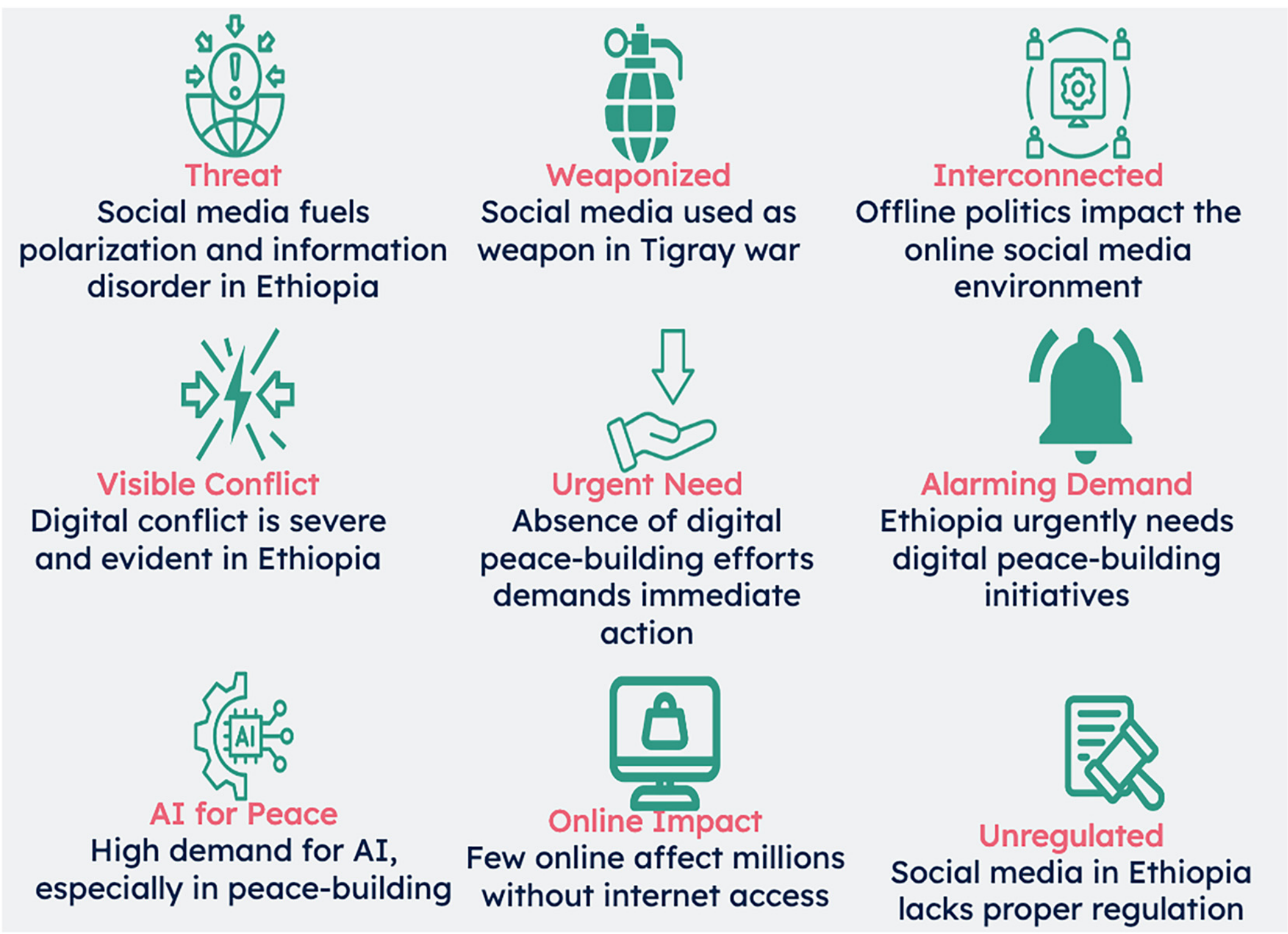

**Figure 4:** The digital media environment in Ethiopia: key issues including polarization, weaponization, digital conflict, offline influence, urgent peace-building needs, widespread impact on unconnected populations, and regulation gaps. Features of the social media environment in Ethiopia: polarization, weaponization, digital conflict, offline influence, urgent peacebuiding needs, and regulation gaps.

Our research findings are summarized in Figure 4 which shows the relationship between social media, polarization, and conflict escalation as well as the absence of digital peace building works. To sum up, the findings reveal that social media plays a detrimental role in the Northern Ethiopia conflict by exacerbating polarization and marginalizing women's voices. It has been weaponized to spread harmful narratives and incite violence, further entrenching divisions. Women's participation in peacemaking efforts is significantly absent, resulting in their perspectives being excluded from important discussions. The intersection of gender, socio-economic status, and ethnicity amplifies the challenges women face in effectively utilizing digital media. The digital divide, government control, and a lack of organized digital peace initiatives further restrict women's use of ICTs, preventing them from leveraging social media to foster engagement and promote peace.

This section highlights how social media contributes to conflict escalation in Northern Ethiopia through processes of polarization, underpinned by social identity, feminist, and intersectionality theories. The findings reveal that online narratives reinforce group loyalties, deepen divisions, and fuel conflict. The analysis underscores the importance of integrating digital peacebuilding tools, addressing structural barriers like the digital divide, and fostering inclusive, gender-sensitive approaches.

## 6 Conclusion and implications

In conclusion, this study underscores the complex influence of social media on conflict during the Northern Ethiopia War. The findings reveal that digital platforms, while holding potential for peacebuilding, have predominantly been wielded as tools of polarization, misinformation, and violence escalation. Social media has amplified ethnic tensions and been weaponized. These consequences extend beyond digital spaces, severely affecting millions, particularly women, who faced life-threatening situations, displacement, and gender-based violence, including mass rapes. Despite women's inherent capabilities for peacebuilding, they continued to be marginalized in the peacemaking process, limiting their positive impact on stability and reconciliation. Furthermore, significant structural barriers including the digital divide, government shutdown, and limited digital literacy, etc. hamper the development of effective digital peacebuilding strategies.

This qualitative study advanced our understanding of the impact of social media on women in violent conflicts. We have explored the dual role of digital technologies in conflict, peacebuilding and gender issues. Through empirical insights from the understudied Northern Ethiopia War, we addressed a gap in research on digital technology's impact in civil strife contexts. This highlights the gap in information system (IS) research regarding the area of tension between gender and technology in conflict scenarios. Therefore, the implications of the study extended to theory and practice. First, it advanced the use of social identity theory in the context of social networks by illustrating how social media can both express and amplify ethnic hostilities, thereby driving polarization. Furthermore, it contributed to liberal feminist theory by emphasizing the need for women's equal participation in peacebuilding and spotlighting the structural barriers they face in conflict-affected regions. Lastly, the study contributed to the intersectionality theory by demonstrating how overlapping identities influence women's engagement and participation in peace initiatives, providing valuable insights into the complexities of marginalization and empowerment.

To address the significant issues identified about social media polarization and weaponization, women's marginalization and challenges of digital peacebuilding in Ethiopia, a multifaceted approach is necessary. First, promoting peace through genuine dialogue and democratic engagement is crucial for transforming the polarized social media environment and fostering a culture of mutual understanding. Furthermore, enhancing digital literacy, especially among women, is of paramount importance. This will empower them to safely and effectively navigate social media platforms, facilitating their active engagement in online discussions and reducing the digital divide. Advancing digital peacebuilding initiatives tailored to leverage the positive aspects of social media can improve communication and cooperation during conflicts, thus contributing to IS design orientation. Additionally, deploying AI technologies presents a valuable opportunity to analyze social media interactions and identify harmful trends, thereby creating a roadmap for peacebuilding activities. Integrating women's voices through inclusive peacebuilding strategies is also essential. Initiatives that actively involve women in the process will help reduce marginalization and enhance their contributions to sustainable and inclusive peace. Engaging community leaders and influencers to challenge societal norms around gender can further encourage women's participation in peacebuilding initiatives. Lastly, implementing proper regulations for social media platforms is critical, as accountability measures will help ensure responsible use of digital communication, thereby improving the overall environment for conflict resolution. These recommendations will assist Ethiopia and other countries in similar contexts in addressing their sociopolitical challenges and working toward sustainable peace.

**Limitation of the study and recommendations for future research**: The study comes with limitations. First, it does not include the direct experiences of war victims or combatants, and therefore does not capture their perspectives firsthand. Instead, the insights are based on accounts from experts working with these groups. Second, the findings may be influenced by the subjective perceptions of participants, who may have biases related to their impressions of social media's role and the effects of digital echo chambers. These subjective perspectives could have affected the experiences reported and, consequently, the research outcomes. Although we sought to mitigate this by triangulating data from multiple sources including interviews, focus group discussions, and document analysis these factors may still limit the generalizability of the results. We acknowledge these limitations and suggest that future research incorporate direct perspectives from war victims and combatants, as well as quantitative measures to address potential biases.

Further research could also involve analyzing social media content produced during the war to determine the level and type of social media polarization. Utilizing network analysis could help to explore how algorithmic bias and echo chambers contribute to polarization, as well as to understand the structure, relationships, and dynamics within social media networks. In addition, employing user-centered approaches, such as participatory design, ethnographic methods, or user experience research, would allow for a deeper understanding of user needs, perceptions, and engagement strategies in digital peacebuilding. Comparative studies examining similar conflicts in other regions or contexts would enable cross-cultural and cross-platform analyses of digital polarization and peacebuilding tools, thereby informing the development of more inclusive and effective peacebuilding strategies.

**Acknowledgments:** The research was partially funded by the Google Award for Inclusive AI Research. Further information on the project can be found here https://www.hcds.uni-hamburg.de/en/research/all-projects/ai-map.html.
**Research ethics:** Not applicable.
**Informed consent:** Informed consent was obtained from all individuals included in this study. Due to the conflict situations of the study area and the sensitivity of the research topic all consents are given orally.

**Author contributions:** All authors have accepted responsibility for the entire content of this manuscript and approved its submission.

**Use of Large Language Models, AI and Machine Learning Tools:** An LLM was used to improve the language.

**Conflict of interest:** The authors state no conflict of interest.

**Research funding:** The research is funded from the Google Award for Inclusive AI Research.

**Data availability:** The raw data can be obtained on request from the corresponding author.

# Appendix A: Coding and data structure

**Table A.1:** Thematic analysis of the data – open code.

| ID | Open code |
|---|---|
| O1 | Women and children were more affected than men during conflict |
| O2 | Mass killings |
| O3 | Sexual violence and rape incidents during the war |
| O4 | Mass displacement |
| O5 | Human slaughtering |
| O6 | Underreporting of sexual violence due to social discrimination and safety concerns |
| O7 | Schools and hosptitals were damaged by the war |
| O8 | Lack of digital platforms for secure reporting of sexual violence |
| O9 | The Tigray war |
| O10 | Gender based violence |
| O11 | Displacement of women and children |
| O12 | The Northern Ethiopia conflict |
| O13 | Women involved in traditional social support systems (Idir, Equb, Coffee ceremonies) |
| O14 | Bloody war |
| O15 | Women's participation in peace processes is minimal or symbolic |
| O16 | Physical violence |
| O17 | Media as weapon of war |
| O18 | Women excluded from peace negotiations and agreements |
| O19 | Psychological trauma |
| O20 | Cultural and societal norms marginalize women's participation in building peace |
| O21 | War was preached as religious sermon |
| O22 | Limited attempts to involve women through peace conferences and UN initiatives |
| O23 | Gang rape as war crime |
| O24 | Limited attempt to involve women in peace ambassadors' role |
| O25 | Social discrimination and marginalization |
| O26 | The hostile online environments |
| O27 | Social media as a weapon for propaganda, misinformation, and hate speech |
| O28 | Social media posts propagating violence |
| O29 | Divisive language use |
| O30 | Genociders |
| O31 | Kill and arrest all Tigrians |
| O32 | Call TPLF as junta, and day-hyena |
| O33 | Call the Ethiopian government as fascist |
| O34 | Labeling the Ethiopian army and the allied forces as genociders |
| O35 | Ethiopian women historical positions |
| O36 | Misinformation, disinformation, and hate speech increasing social polarization |
| O37 | Use of Facebook, TikTok, Instagram fueling ethnic nationalism and violence |
| O38 | Social media as tool in the armed struggles |
| O39 | Hate speech laws have gaps and enforcement is weak |
| O40 | Social media platforms used to spread inflammatory music and messages (e.g., Afan Oromo) |

**Table A.1:** (continued).

| ID | Open code |
|---|---|
| O41 | Misinformation becomes primary driver of offline conflict |
| O42 | Social media campaigns impacting public perception and fuels ethnic tensions |
| O43 | The polarized media environment distorted the reality of the war |
| O44 | Low media literacy skills exacerbate polarization and disinformation |
| O45 | Limited media literacy programs and fact-checking efforts underway (e.g., Arts TV, EBC) |
| O46 | Digital peacebuilding efforts are non-existent or limited |
| O47 | No comprehensive digital peace strategies or frameworks |
| O48 | NGOs and CSOs implementing digital initiatives for GBV and peace advocacy with limited scope |
| O49 | Unsatisfactory use of chat groups, social media channels, and hot-line for peace messaging and GBV support |
| O50 | Lack of government-led digital peacebuilding initiatives |
| O51 | Mistrust between citizens and government authorities |
| O52 | Traditional community forums serve as spaces for social cohesion |
| O53 | Political reforms required to address root causes of conflict |
| O54 | Ethnic nationalism and identity politics as persistent online conflict drivers |
| O55 | Elite rivalry and competition over political and economic power |
| O56 | External regional influences (e.g., Eritrea) exacerbating conflict |
| O57 | Political economy issues fueling conflict |
| O58 | Poor professionalism and bias in government media |
| O59 | Self-censorship and influence of political parties in media |
| O60 | Societal norms reinforce patriarchal exclusion of women |
| O61 | Women's economic and social empowerment constrained by societal norms |
| O62 | Women's organizations utilize digital platforms to combat GBV is weak |
| O63 | Digital literacy among the population is low |
| O64 | ICT infrastructure is not adequately developed |
| O65 | Media polarization worsened by political discourse and illiteracy |
| O66 | NGOs working for gender empowerment and peace advocacy |
| O67 | Need for legal reforms on hate speech and disinformation |
| O68 | Political systems and institutions lack inclusivity and transparency |
| O69 | The media environment characterized by bias, misinformation, and low professionalism |
| O70 | Societal trauma resulting from war and digital conflicts |
| O71 | Social media aggravates historical grievances |
| O72 | Attempts at peace dialogues via social media platforms and civil society groups |
| O73 | The Pretoria Peace Agreement |
| O74 | The importance of political commitment for peacebuilding success |
| O75 | Digital divide |
| O76 | Online safety and security |
| O77 | Human right violation |
| O78 | Digital conflict in the absence of digital peacebuilding |
| O79 | Government shutdown of the internet and its control |
| O80 | Digital tools (AI, sentiment analysis) are underutilized but promising |
| O81 | The role of digital media in the peacebuilding process is almost zero |
| O82 | Social media is the major problem of the nation |
| O83 | Capacity limitation |
| O84 | Absence of synergy for peacebuilding |
| O85 | Lack of political commitments |
| O86 | Amhara region |
| O87 | Afar region |
| O88 | Ethiopia's complex ethnic fabric and political tensions often spill into the digital space |
| O89 | The digital conflict, while having limited coverage, manifests widespread offline consequence |
| O90 | 'us' versus 'them' |
| O91 | Devil |
| O92 | Tigray People's Liberation Front (TPLF) |
| O93 | Amhara National Movement (ANM) |
| O94 | Oromo Liberation Front (OLF) |

**Table A.2:** Thematic analysis of the data – axial code.

| ID | Axial code |
|---|---|
| A1 | Use of social media for propaganda, misinformation, and hate speech that exacerbates violence |
| A2 | Amplification of gender-based violence (GBV) including sexual violence |
| A3 | Impact of misinformation on women's safety, societal perceptions, and psychological well-being |
| A4 | Limited digital spaces for women's voices; marginalized women's participation and influence |
| A5 | Limited use of digital platforms (Telegram, Facebook, YouTube, X) for advocacy (GBV support, peace dialogues) |
| A6 | Women's exclusion from peace negotiations and agreements (e.g., Pretoria Peace Agreement) |
| A7 | Need for gender-sensitive approaches and institutional reforms to ensure women's meaningful participation |
| A8 | Societal (patriarchal and cultural) norms hindering women's political and peace participation |
| A9 | Efforts (though limited) to increase women's involvement (e.g., women peace ambassadors, NGOs) |
| A10 | The gap between policy aspirations and actual women's inclusion and participation |
| A11 | Social media used to fuel ethnic nationalism, division, and armed conflict |
| A12 | Misinformation, disinformation, and hate speech as tools for ethnic tension and political escalation |
| A13 | Social media platforms (Facebook, TikTok, Instagram, YouTube and X) amplifying stereotypes, hostility, and inflammatory content |
| A14 | Political interests reinforcing polarization |
| A15 | Social media as a battleground for ethnic identity politics |
| A16 | The potential of social media for peacebuilding is unused |
| A17 | Absence of government-led or institutionalized digital peace strategies |
| A18 | Absence of digital peace buildings |
| A19 | Recognition among stakeholders about the need for structured digital peace initiatives |
| A20 | Weak legal frameworks and gaps in hate speech and disinformation laws |
| A21 | Limited technical capacity, infrastructure, and resources for advanced digital peace tools (AI, analytics) |
| A22 | Lack of coordination among government, CSOs, and technology actors |
| A23 | Media illiteracy, misinformation proliferation, and unprofessionalism of media outlets |
| A24 | Ensuring sustainability and inclusivity of digital peace efforts remains a major challenge |
| A25 | CSOs and NGOs deploying digital tools (chat groups, social media campaigns, hot-lines) but with limited scale |

**Table A.3:** Thematic analysis of the data – selective theme.

| No. | Selective theme |
|---|---|
| T1 | The role of social media in exacerbating conflict and gender-based violence during the Northern Ethiopia War |
| T2 | The persistent marginalization of women in peacemaking and peacebuilding processes |
| T3 | The polarization and weaponization of social media in the Ethiopian context |
| T4 | The dearth of digital peacebuilding initiatives in Ethiopia |
| T5 | The challenges of digital peacebuilding in Ethiopia |

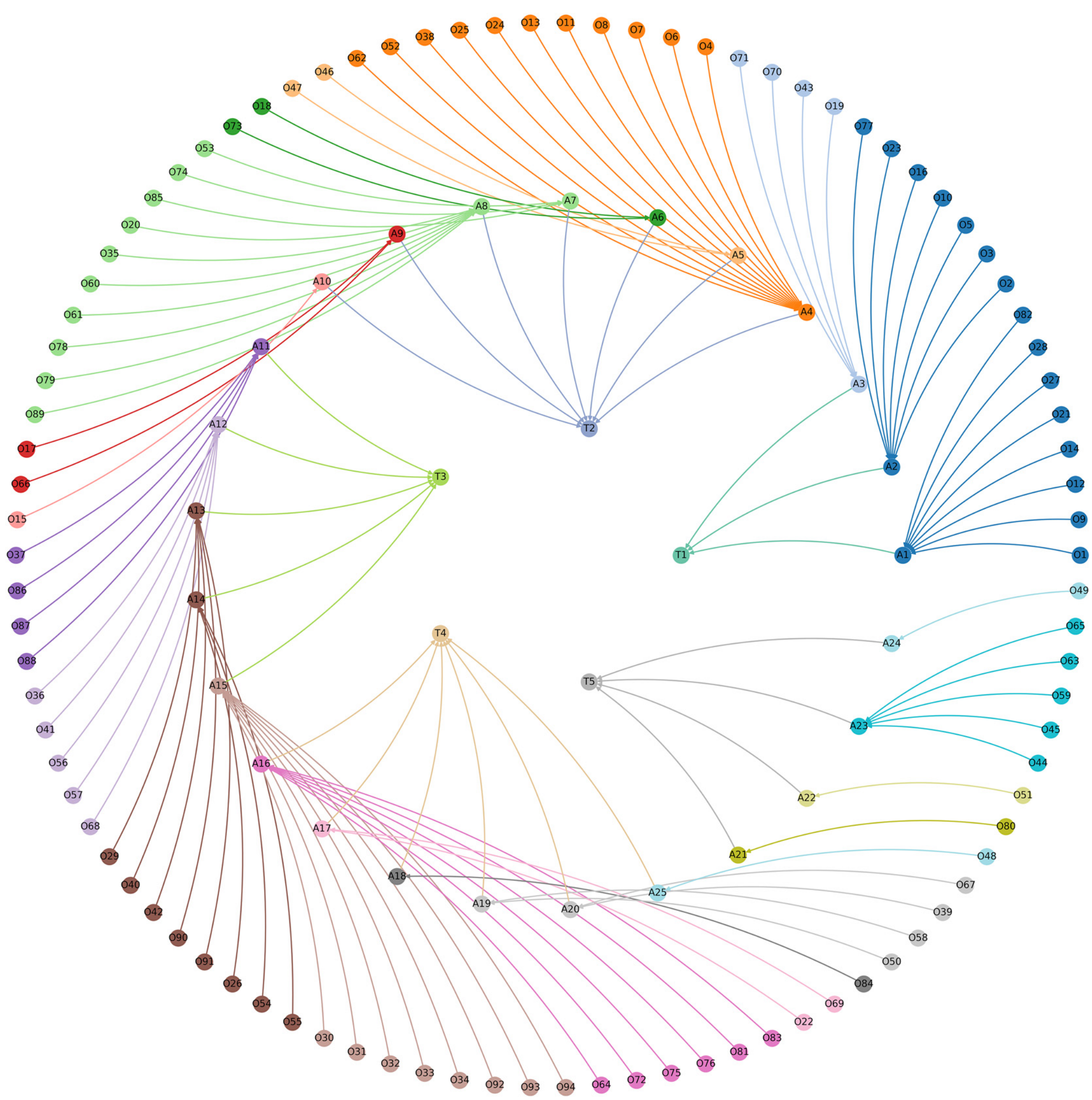

**Figure A.1:** Visual representation of the coding scheme: open codes, axial codes, and selective themes with color-coded groups.